\def \frac#1#2{ { #1 \over #2} }
\def \R {\phi}
\newcommand{\refp}[1]{(\ref{#1})}
\newcommand{\beq}{\begin{equation}}
\newcommand{\beqar}{\begin{eqnarray}}
\newcommand{\eeq}[1]{\label{#1}\end{equation}}
\newcommand{\eeqar}[1]{\label{#1} \end {eqnarray}}

\documentstyle[preprint,aps]{revtex}

\begin{document}

\title{Semiclassical Approximation to Neutron Star Superfluidity 
Corrected for Proximity Effects}

\author{ F. Barranco$^{a}$, R.A. Broglia$^{b,c}$,
        H. Esbensen$^{d}$ and E. Vigezzi$^{b}$}            

\address
{$^a$Dpto. de F\'\i sica Aplicada, Escuela Superior de Ingenieros,
   Universidad de Sevilla, Spain. }

\address
{$^b$Dipartimento di Fisica, Universit\`a di Milano and 
INFN, Sezione di Milano, Italy.}

\address
{$^c$The Niels Bohr Institute, University of Copenhagen, Denmark.}

\address
{$^d$Physics Division, Argonne National Laboratory, Argonne, 
Illinois 60439, USA. }
 
\date{\today}

\maketitle

\vskip 1cm

\begin{abstract}

The inner crust of a neutron star is a superfluid and inhomogeneous 
system, consisting of a lattice of nuclei immersed in a sea of neutrons.
We perform a quantum calculation of the associated pairing gap and 
compare it to the results one obtains in the Local Density 
Approximation (LDA). It is found that the LDA overestimates the spatial 
dependence of the gap, and leads to a specific heat
of the system which is too large at low temperatures, as compared 
with the quantal result. This is caused by 
the neglect of proximity effects and the delocalized character
of the single-particle wavefunctions close to the Fermi energy. 
It is possible to introduce an alternative, simple semiclassical 
approximation of the pairing gap which leads to a specific heat that 
is in good agreement with the quantum calculation.

\end{abstract}



keywords: pairing gap, neutron stars

\newpage

\section{Introduction}

Above a density of about  about 3 $\times$ 10$^{11}$ g cm$^{-3}$
(corresponding to 1.8 $\times 10^{-4}$ n fm$^{-3}$)
matter in neutron 
stars consists of a matrix of nuclei immersed in a sea of neutrons and
an approximately uniform sea of electrons. Such a configuration persists
up to roughly half nuclear saturation density, and constitutes the 
so-called "inner crust" of a neutron star. 
At low temperatures, this system is 
superfluid with a positive Fermi energy $\epsilon_F$, 
and one is confronted with the task of estimating the pairing 
gap $\Delta$ of nuclei immersed in the neutron liquid \cite{Pines} . 
With the exception of ref. \cite{noi},
the pairing gap in this inhomogeneous
medium has been calculated assuming that locally it coincides 
with the value of the pairing gap 
in bulk matter at the corresponding density 
(cf. e.g. ref. \cite{Pethick} and refs. therein). 

A  complete calculation of pairing should include 
the induced interaction, taking into account polarization effects in
the medium (cf. for example \cite{Pines}).
 In the present paper we will restrict ourselves to the mean field level,
and we shall compare the results of 
a local density approximation (LDA) calculation of
the pairing gap to the results obtained from a Hartree-Fock-Bogoliubov (HFB) 
calculation. We find that the LDA leads to a spatial variation of the
gap near the surface of a nucleus which is stronger
than that obtained in the HFB calculation.
This is because a Cooper pair in the quantum (HFB) calculation samples 
the pairing interaction over a relatively large distance, which is of the 
order of the coherence length  (cf. Section III).
As a result, the pairing
interaction in the neutron liquid influences the gap in nuclei, and 
vice versa, so that the spatial dependence of the gap is smeared out
near the nuclear surface.

The pairing gap plays an important role in the thermal evolution of a 
neutron star, because it influences in an essential way the specific 
heat in the inner crust.
We will show that the LDA overestimates the effect of the presence of 
nuclei on the specific heat by a large amount (cf. Section IV).
In the LDA, the specific heat is obtained from an integral in phase
space, and it is possible to single out two contributions to the 
specific heat, 
one from the interior of the nucleus and one from the outer neutron liquid. 
Inside the nucleus the local pairing gap is small, and this enhances
the value of the specific heat. 
On the other hand, in the quantal  calculation the specific heat 
is determined from  the value of the pairing gap of the  states close
to the Fermi surface, which is obtained from an integral over the whole
Wigner cell, and is much less sensitive to the presence of the nucleus.

\newpage
\section{Pairing field in coordinate space}

We consider a spherically symmetric system which is governed by a
single-particle Hamiltonian {\it h} and a two-body interaction {\it v}.
The Hamiltonian {\it h = T + U} consists of the kinetic energy and a mean-field
$U$ which we do not calculate self-consistently but parametrize
as described later on. The explicit effect of the two-body interaction 
on the ground state of the system is therefore included in the HFB
equations only via the pairing field.

The system is placed into a spherical box with a nucleus at the origin.
The radius of the box is equal to that of the Wigner-Seitz cell of 
the lattice (cf. Fig. 1). Nuclei belonging to different cells 
are so far away from each other, that 
one can solve for a single cell separately.
The single-particle wavefunctions, which are eigenfunctions to {\it h}
are therefore required to vanish at the boundary of the cell.
They are denoted by $\phi_{nljm} (\vec r) $ = $\vec r|nljm>$, while 
$\epsilon_{nlj}$ are the associated eigenvalues.

The HFB equations \cite{Ring} are solved in matrix form using this 
basis. We shall, in particular, make use of the two-particle 
wavefunctions
\beq
<\vec r_1\vec r_2|nn'lj,J=0> \ = \R_{nlj}(r_1)
\R_{n'lj}(r_2)
\sum_{m} {(-1)^{l+j-m}\sqrt{2}\over 2j+1} 
<{\hat r}_1|ljm> <{\hat r}_2|lj-m>,
\eeq{2pwf}
where $J$ is the total 
angular momentum of the system and $\phi_{nlj}$ is a radial wavefunction.
The normalization of
these two-particle wavefunctions is $2/(2j+1)$.

In the case of spherical symmetry, 
the HFB equations are diagonal in the quantum numbers {\it (lj)}, 
having the form
\beq
(\epsilon_{nlj}-\epsilon_F) U^q_{nlj} 
+\sum_{n'} \Delta_{nn'lj} V^q_{n'lj} = E_{qlj} U^q_{nlj},
\eeq{HFB1}
\beq
-(\epsilon_{nlj}-\epsilon_F) V^q_{nlj} 
+\sum_{n'} \Delta_{nn'lj} U^q_{n'lj} = E_{qlj} V^q_{nlj},
\eeq{HFB2}
where $\epsilon_F$ is the Fermi energy.
The eigenvalues $E_{qlj}$ of these equations are the energies of
the quasiparticle states, while the $U$ and $V$ eigenvectors are the 
occupation amplitudes on the 
single-particle states.

The matrix elements of the pairing field, $\Delta_{nn'lj}$,
can be expressed as
\beq
\Delta_{nn'lj} = - <nn'lj; J=0|v|\Phi>,
\eeq{HFB}
where 
\beq
|\Phi> = \sum_{nn'lj} \frac{(2j+1)}{2}
\Bigl(\sum_q U^q_{nlj} V^q_{n'lj}\Bigr) |nn'lj; J=0>
\eeq{Cooper}
is the pairing density, which is related to the abnormal density
as discussed in ref. \cite{Dobaczewski}. 

The HFB equations are solved self-consistently.
Each set of solutions $(U,V)$ determines a new pair density $\Phi$ 
and new matrix elements of the pairing field, which are then used
in the next iteration. This scheme is continued
until a converged set of solutions has been obtained.
The two particles in Eq. \refp{2pwf} can be in states with a 
different number of nodes, while in the BCS approximation they
are assumed to be in time reversed states ($n=n'$).
The condition $n=n'$ is in general too restrictive for an accurate
description of localization effects and of the coupling of bound and 
continuum states \cite{noi,Dobaczewski,Nazarewicz}.

\subsection{S=0 pairing}

We shall carry out the calculation making use of a Gogny force \cite{Gogny}.
In this case, 
the strongest part of the interaction $v$ is the attractive $S$=0
component.
The repulsive $S$=1 component is much weaker and, within the present context, 
we shall ignore it.
The spatial part of the $S$=0 component of the two-particle 
wavefunction \refp{2pwf} is
\beq
\psi_{nn'lj}(\vec r_1,\vec r_2) =
<\vec r_1,\vec r_2|nn'lj;J=0>_{S=0}=
{1\over 4\pi} \ 
\R_{nlj}(r_1) \R_{n'lj}(r_2) \ 
P_l(cos(\theta_{21})), 
\eeq{rho}
where $\theta_{21}$ is the angle between $\vec r_1 $ and $\vec r_2$.
Inserting this expression into the $(\vec r_1 , \vec r_2) $ 
representation of the function defined in Eq. \refp{Cooper} we obtain the 
pairing density
\beq
\Phi_{S=0}(\vec r_1,\vec r_2) = \sum_{nn'lj}
\Bigl(\sum_q U^q_{nlj} V^q_{n'lj}\Bigr)
{2j+1\over 8\pi} \ 
\R_{nlj}(r_1)\R_{n'lj}(r_2) \
P_l(cos(\theta_{21})).
\eeq{Cooper1}
This quantity 
is symmetric in the positions of the two particles, whereas the
spin part, $|S=0>$, is antisymmetric. Equation \refp{Cooper1}
describes a correlated two-particle state, the so-called Cooper pair
wavefunction. Because of the $U,V$ factors, this wavefunction receives
contributions essentially only from single-particle states around the Fermi
surface.

The pairing field is defined as product
$\Delta(\vec r_1, \vec r_2)= -v(|\vec r_1-\vec r_2|)
\Phi_{S=0}(\vec r_1, \vec r_2)$.
The pairing matrix elements are calculated,
(cf. Eq. \refp{HFB}), by projecting this field onto the 
two-particle wavefunctions \refp{rho}.
For later discussion we introduce the Fourier transforms of the
pairing field $\Delta$ and of the two-particle wavefunctions defined in 
Eq.\refp{rho},
\beq
\Delta (\vec k, \vec R) = \int d^3 r_{12} 
\ \Delta (\vec r_1 , \vec r_2) \
e^{-i \vec k \cdot \vec r_{12}} 
\eeq{Ft1}
\beq
\psi_{nn'lj}(\vec k,\vec R) = \int d^3 r_{12}
\ \psi_{nn'lj}(\vec r_1,\vec r_2) e^{-i\vec k \cdot \vec r_{12}},
\eeq{Ft2}
where $\vec r_{12}$ is the relative distance of the two neutrons and 
$\vec R$ is the center of mass of the pair.
The pairing matrix elements can then be written
\beq
\Delta_{nn'lj} = {1\over (2\pi)^3}
\int d^3 k \int d^3 R 
\ \Delta(\vec k,\vec R) \ \psi_{nn'lj}(\vec k,\vec R).
\eeq{connft}

\subsection{Local Density Approximation}

Setting the nuclear potential $U$ equal
to zero, and allowing the radius $R_b$ of the Wigner-Seitz cell go to
infinity for a given value of the Fermi energy $\epsilon_F$, 
one recovers the results valid  for uniform neutron matter.
In fact, for  $R_b \to \infty$ one can substitute  $\R_{nlj}$
by $(2k^2/R_b)^{1/2} j_l(kr)$, 
and the sums over $n,n'$ by $(R_b/\pi)^2 \int dk \int dk'$.
In this limit, the matrix element of the gap defined in Eq.\refp{HFB} 
becomes diagonal in $k$ and $k'$, and the HFB equations reduce to 
those of the BCS theory. The pairing density of Eq. (5) has only one spin 
component in this case, namely the $S=0$, and it is given by
\beq
\Phi_{unif}(r_{12}) = 
\int \frac {d^3 k} {(2 \pi)^3} U_k V_k e^{i \vec k \cdot \vec r_{12}}.
\eeq{unif}
As expected, the Cooper-pair wavefunction depends, in this limit, only 
on the relative distance $r_{12}$  of the two neutrons.
This is also the case for the associated pairing field
\beq
\Delta_{unif}(r_{12}) = - v(r_{12})
\int \frac {d^3 k} {(2 \pi)^3} U_k V_k e^{i\vec k \cdot \vec r_{12}}.
\eeq{unifd}
The Fourier transform of this quantity is 
\beq
\Delta_{unif}(k, \epsilon_F) = 
- \int \frac {d^3 k'} {(2 \pi)^3} 
\frac {\Delta_{unif}(k',\epsilon_F)} {2 E_{k'}} 
v(| \vec k - \vec k'|). 
\eeq{uniff}
In this expression 
we have explicitly indicated the dependence of the pairing gap 
on the Fermi energy, and 
the BCS relation $U_k V_k = \Delta(k)/2 E_k$ has been used, the
quasiparticle energy 
$E_k$ 
being
\beq
E_k = \sqrt { ( {\hbar^2 k^2}/{2m} - \epsilon_F)^2 + 
\Delta^2_{unif} (k, \epsilon_F)}.
\eeq{Ek}

The LDA to the pairing gap is obtained (cf. refs. 
\cite{Bengtsson,Winther}; cf. also \cite{Oliveira})
by replacing $\epsilon_F$ in Eqs. \refp{uniff} and \refp{Ek}
by the local Fermi energy  $ \epsilon_F(R) = \epsilon_F - U(R)$.
The corresponding local Fermi momentum is 
given by $\hbar^2 k_F^2(R) = 2 m \epsilon_F(R)$. 
In this way one obtains a space-dependent
pairing gap $\Delta_{LDA} (k,R)= \Delta_{unif} (k, \epsilon_F(R))$ 
and a space dependent quasiparticle energy,
\beq
E_k(R) = \sqrt { ( {\hbar^2 k^2}/{2m} - \epsilon_F(R))^2 
+ \Delta^2_{unif}(k,\epsilon_F(R))}.
\eeq{Ekr} 
First- and second-order corrections in powers of $\hbar$ to the 
LDA have also been studied \cite{Taruishi}.

\section{Comparison of HFB and LDA results}

In the following we present 
calculations of the quantum and LDA gaps 
$\Delta(\vec k, \vec R)$ for a
Wigner-Seitz cell of 29 fm, and a Fermi energy $\epsilon_F =13.5$ MeV
corresponding to $^{1800}_{50}$Sn and typical of the situation  
encountered at the density 
of about  3  $\times 10^{13}$ g cm $^{-3}$ ( or $\rho \approx$ 0.02 n fm$^{-3}$)
in the "inner crust" of a neutron 
star. 
As for the interaction $v$, we adopt the Gogny force in the
$^1S_0$ channel \cite{Gogny}. 
In the HFB calculations, we have performed the self-consistent
calculation only for the pairing field, while instead of 
the Hartree-Fock  field
we have used a Woods-Saxon potential, without introducing an 
effective mass.
The adopted Wood-Saxon potential (cf. Fig. 1) 
has a depth $V_o $ = -31 MeV, a radius  $R =$ 7.5 fm  and a diffusivity  
$a = $ 0.9 fm. These values 
are chosen in order to reproduce 
the neutron density
in the cell, calculated in ref.  \cite{Negele}.
The resulting neutron density is equal
to about $\rho_{int}$= 0.1 n fm$^{-3}$ in the 
interior of the nucleus (R $<$ 5 fm), 
and then drops to about $\rho_{ext}= 0.018$ n fm$^{-3}$ for 
11 fm $< R < $ 27 fm,
until it goes to zero close to the boundary of the cell. 
We have included the single-particle levels up to 100 MeV, controlling
the
convergence of the results. We have also checked that in the absence of
the Woods-Saxon potential, we obtain the pairing gap and specific heat
of neutron matter at the same Fermi energy. The pairing gap in neutron matter
$\Delta_{unif} (k,\epsilon_F)$, 
calculated at  $k= k_F$, is shown in Fig. 2 
as a function of $\epsilon_F$. In Fig. 2 we also show the value of 
the local Fermi energy at three points in the Wigner cell.

In neutron matter one can write $\Phi_{unif} \sim$ sin$(k_F r_{12})/k_F r_{12}$
as one can see from Eq. \refp{unif} taking into account
the fact that the product $U_k V_k$ is peaked at the Fermi energy,
so that it oscillates with 
a wavelength given by 
$\lambda \approx 2 \pi/k_F$. These oscillations are damped over a scale 
which is approximately determined by the coherence length 
$\xi = \hbar^2 k_F/m \pi \Delta$
\cite{De Gennes,Weisskopf}. 
The pairing gap calculated with 
the Gogny interaction in neutron matter at the Fermi energy $\epsilon_F$ 
=13.5 MeV 
is about 3.5 MeV (cf. Fig. 2), so that one obtain 
$\xi =$ 3.2 fm. 

In Fig. 3  we show instead the square of the Cooper pair 
wavefunctions calculated with HFB in the presence of the nucleus 
at three  values of $R$ ($R$= 3 fm, 8 fm and 15 fm), as a function of $r_{12}$.
At large values of $R$, well outside the nucleus (cf. Fig. 3(c)), 
the wavefunction is essentially the same 
as calculated in uniform neutron matter 
at the same Fermi energy, $\epsilon_F= 13.5$
MeV. The radius of the Cooper pair,  given by the 
mean square relative distance, 
$<r_{12}^2>^{1/2}$,  is about 4 fm, to
be compared with the coherence length of 3.2 MeV reported above.
Inside the nucleus, at $R=$ 3 fm (cf. Fig. 3(a)), or on the surface, 
at $R$= 8 fm (cf. Fig. 3(b)), the wavefunction feels the presence of
the nucleus, 
but its main features 
can still be understood in the same way, using the local values of the Fermi
energy and of the pairing field.
In this case the wavefunction shows some dependence also on the angle  
$\theta$ between $\vec R$ and $\vec r_{12}$
(especially for large values of $r_{12}$), and we show 
the results for $\theta=45^o$.
The local Fermi energy decreases going from inside to outside the nucleus
(cf. Fig. 2):
it is equal to about
44.5 MeV at $R=$ 3 fm,
instead of about 25 MeV at $R=$ 8 fm or 13.5 MeV at $R$= 15 fm.
Therefore the local wavelength $2 \pi / k_F(R)$
of the Cooper pair  is much shorter inside than outside  the nucleus, 
as can be seen comparing Fig. 3(a) with Fig. 3(b) or Fig. 3(c).
Moreover, the pairing gap 
$\Delta_{LDA}(k,\epsilon_F(R))$ calculated
at the local Fermi momentum $k_F(R)$, denoted by  $\Delta(k_F(R))$, 
is much smaller inside the nucleus than on the surface or outside,
(cf. Fig. 2 and Fig. 4(b)), so that 
the coherence length is larger at
$R=$ 3 fm, because $k_F$ is larger, and $\Delta$ is smaller.
Using the values of $\Delta(k_F(R))$ 
(1.6  MeV at $R=$ 3 fm and 3.5 MeV at $R=$ 8 fm, cf. Fig. 2), 
one obtains $\xi =$ 12  fm  at $R=$ 3 fm and $\xi =$4.1 fm
at $R=$ 8 fm. These  values can be compared to the calculated r.m.s.
values of 10.3 fm (at $R=$ 3 fm) and of 3.8 fm (at $R=$ 8 fm).

The pairing gaps $\Delta(k,R)$ and $\Delta_{LDA}(k,R)$ are presented 
in Fig. 4(a) and Fig. 4(b) for different values of $k$ as a function of $R$.
The value of $\Delta (\vec k ,\vec R)$ depends on the angle between 
$\vec k$ and $\vec R$, but we have found that this dependence is 
very small in the present case.
The overall dependence on $k$ and $R$ is similar in the two 
calculations, but the HFB gap shows a smoother spatial variation than 
the LDA result. This is due to the rather large extension of the Cooper
pairs, as discussed above and shown in Fig. 3, leading to important 
proximity effects, so that the difference in the gap between
the interior of the nucleus and the outer part is smeared out.
In the semiclassical case, we also show the values of the gap calculated at
the local Fermi momentum as a function of $R$. It is seen that  
$\Delta_{LDA}(k_F(R))$ is equal to about 1.6 MeV and 3.5 MeV inside
and outside the nucleus respectively, as already discussed above.



\section{The specific heat}

A quantity playing an important role in the study of neutron stars, 
is the specific heat of the superfluid phase of the inner crust  
(cf. e.g. refs.  \cite{Pethick},\cite{De Blasio}, \cite{Lazzari}).
This quantity
is defined as
\beq
C_v = \frac {1}{V} \frac {\partial < E >}{\partial T}, 
\eeq{cv}
where $V$ is the volume of the Wigner-Seitz cell. In the framework of our
mean-field theory, the 
energy of the system is simply obtained summing over the quasiparticle states
$q$:
\beq
< E >  = \sum_q n_q E_q, 
\eeq{equa}
the quantity $n_q = (1 + e^{(E_q/T)})^{-1}$
being the occupation number of the quasiparticle state $q$.

The LDA result is obtained \cite{De Blasio} 
taking the derivative of the quantity
\beq
< E >_{LDA}  = \int \frac {d^3k d^3R}{(2 \pi)^3}
n_k(R) E_k(R), 
\eeq{elda}
which is the semiclassical counterpart of Eq. \refp{equa}. In this
equation, the occupation factor $n_k(R)$ is given by 
$n_k(R) = (1 + e^{(E_k(R)/T)})^{-1}$, where $E_k(R)$ is given by 
Eq. \refp{Ekr}. 

In Fig. 5 we compare the specific heat calculated in HFB and in LDA.
The semiclassical approximation grossly overestimates the specific heat
in the presence of the nuclear potential at low temperatures, while it 
tends to the quantum result at high temperatures.

In order to better understand the main source of this discrepancy, 
it is useful to write explicitly the expressions for $C_v$ at low 
temperatures, when $\Delta \gg T$. It is also convenient  to use the 
BCS approximation, which in the present case produces a value for the 
specific heat that is close to the HFB result.
Using the fact that in BCS $n=n'$, one can simply write $\Delta_{nlj}$ and
$\psi_{nlj}$ instead of  $\Delta_{nn'lj}$ and $\psi_{nn'lj}$, 
 and one obtains   \cite{Landau} 
\beq
C_v = \frac {1}{V} \frac {\partial}{\partial T} 
\sum_{nlj} \Delta_{nlj} e^{-(\epsilon_{nlj} - \epsilon_F)^2/2T 
\Delta_{nlj}}
e^{- \Delta_{nlj}/T}.
\eeq{cvqua}

In the LDA we obtain instead for $\Delta \gg T$,
\beq
C_{v, LDA} = \frac{1}{V}\frac{\partial}{\partial T}
\int \frac {d^3 k d^3 R}{(2 \pi)^3}
\Delta_{LDA}(k,R) e^{-(\hbar^2 k^2/2m - \epsilon_F (R))^2 /
(2T \Delta_{LDA} (k,R))} e^{-\Delta_{LDA} (k,R)/T} 
\eeq{cvlda}
where we have used the relation given in Eq. \refp{Ekr}.

For small temperatures, the dominant contributions to the semiclassical 
specific heat come from the regions of the phase space $(k,R)$ 
close to the line
defined by the local Fermi momentum $k_F(R)$. As shown in Fig. 4(b),
there is a clear difference in the values of $\Delta(k_F(R))$ inside and
outside the nucleus, and the integral in Eq. \refp{cvlda} roughly separates 
in two contributions. 
Although the associated volume is small, 
the contribution from the inner region is very large,
because  the local values of the gap are small and are heavily weighted
by the exponential factors. 
If one uses in Eq. \refp{cvlda} the HFB value of 
$\Delta(k,R) $ instead of the semiclassical values, one obtains an improvement
compared the LDA, 
but the  specific heat  calculated in this way is still
far from the quantum result.
Considering instead the quantum expression \refp{cvqua}
for the specific heat,
and going back to Eq. \refp {connft}, we observe that the value of the gap 
for a given state is obtained averaging the value of $\Delta(k, R)$ over
$\psi_{nlj}(k,R) $.
The main contribution to the specific heat comes from states close 
to Fermi energy, which are very delocalized, and are rather close to 
plane waves. On the contrary, Eq. \refp{cvlda} 
uses the local value of
the gap. This suggests to substitute
the  weighting factor 
$e^{-\Delta_{LDA} (k,R)/T} $ in Eq. \refp{cvlda}  by 
$e^{- < \Delta_{LDA} (k,R) >/T}$, where
\beq
<  \Delta_{LDA} (k,R)>  = \int d^3 R' d^3k'
\Delta_{LDA} (k',R') \psi_{\epsilon}(\vec k',R')
\eeq{dlda}
is the average of the semiclassical gap over the density associated 
with a single-particle state $\phi_{\epsilon}$ of energy close  to
$\epsilon= \hbar^2 k^2/2m + U(R)$.
In the present case, it is sensible to approximate 
the wavefunctions close to the (positive) Fermi energy 
by plane waves $e^{i\vec k_L(R')\cdot \vec R'}/\sqrt{V}$, so that 
\beq
\psi_{\epsilon}(\vec k',R') \approx \frac{1}{V} \delta(\vec k' - \vec k_L(R')),
\eeq{psie}
where the local momentum $k_L(R')$ is defined by
\beq
\frac{\hbar^2}{2m} k^2_L (R') + U(R')= \epsilon, 
\eeq{Kl}
and 
\beq
<\Delta_{LDA}(k,R) > = \frac{1}{V} \int d^3R' \Delta_{LDA}(k_L(R'),R'). 
\eeq{fine}
Using the occupation factor $e^{-<\Delta_{LDA}>/T}$ in Eq. \refp{cvlda},
one obtains a specific heat which is very close to the exact result at 
all temperatures, as illustrated in Fig. 5.

We conclude that the local density approximation to the 
pairing phenomenon  overestimates the spatial dependence of the gap,
because it does not take into account the non-locality of the Cooper 
pairs, and the associated proximity effects.
Particular care has to be taken in estimating the thermal occupation 
factors, for which the LDA can lead to large errors at low temperatures.
However, at least in the present case, it is possible to obtain an 
accurate approximation to the quantum HFB result by a simple averaging of
the semiclassical expression of the pairing gap.

\section{Acknowledgments}

This work was supported in part (H.E.) by the U.S. Department of Energy,
Nuclear Physics Division, under contract No. W-31-109-ENG-38.

\vskip 2cm

{\bf Figure captions}

\vskip 1cm

{\bf Fig. 1} 

The Woods-Saxon potential $U(R)$ in the Wigner-Seitz 
cell of radius $R_b=$ 29 fm is shown. 
We also indicate the Fermi energy $\epsilon_F$ =13.5 MeV, and the outer
neutron density $\rho_{ext}$.

\vskip 1cm

{\bf Fig. 2}

The pairing gap $\Delta_{unif}(k,\epsilon_F)$
calculated with the Gogny interaction in uniform neutron  
matter at the Fermi momentum
$k_F$ is shown as a function of the Fermi energy.
Also indicated by the arrows are the values of the local Fermi energy 
$\epsilon_F(R)$ in
the Wigner-Seitz cell at $R$= 3 fm, 8 fm and 15 fm.

\vskip 1cm

{\bf Fig. 3} 

The dependence of the square of the Cooper pair wavefunction $\Phi(R,r_{12})$
on the relative distance $r_{12}$ is shown for three
different values of the center of mass of the Cooper pair in the
Wigner-Seitz cell, $R=$ 3 fm (a), 8 fm (b) and 15 fm (c).

{\bf Fig. 4 }

The pairing gap $\Delta(k,R)$ is calculated as a function of $R$ in the
Wigner cell, using the HFB (a) or the LDA (b)  equations. 
The gap remains constant
for $R > $ 12 fm.
The various curves refer to  different values of $k$, namely
$k=$ 0.25,0.75,1.25,1.75 and 2.5  fm$^{-1}$ going from the top to the bottom
curve. In figure (b), we also show 
the value of the gap calculated at local Fermi momentum $k_F(R)$
(dashed curve).
 
{\bf Fig. 5} 

Specific heat of a Wigner cell containing a nucleus (in units of MeV k$^{-1}$
fm $^{-3}$)  calculated 
according to the  HFB approximation,
to the LDA approximation,
and to the LDA corrected using the averaged pairing gap
$<\Delta_{LDA}>$
according to Eq. \refp{fine}.

\vskip 2cm
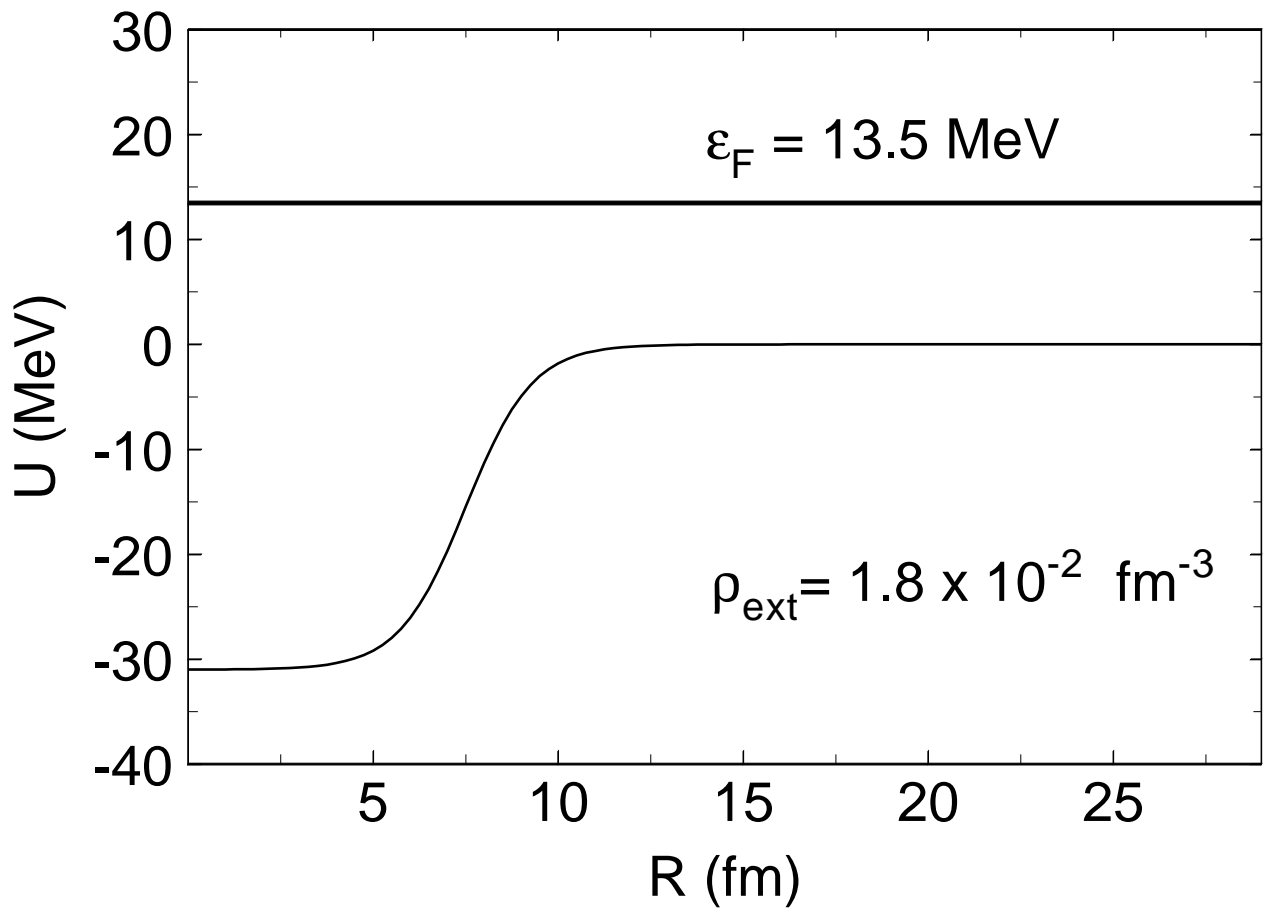
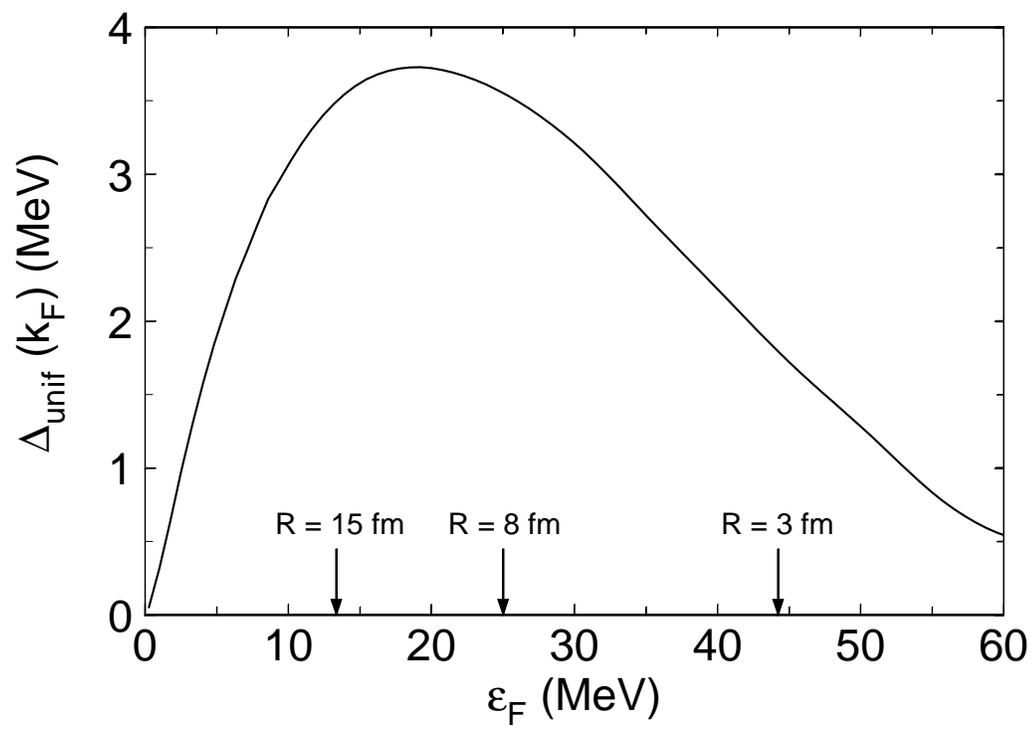
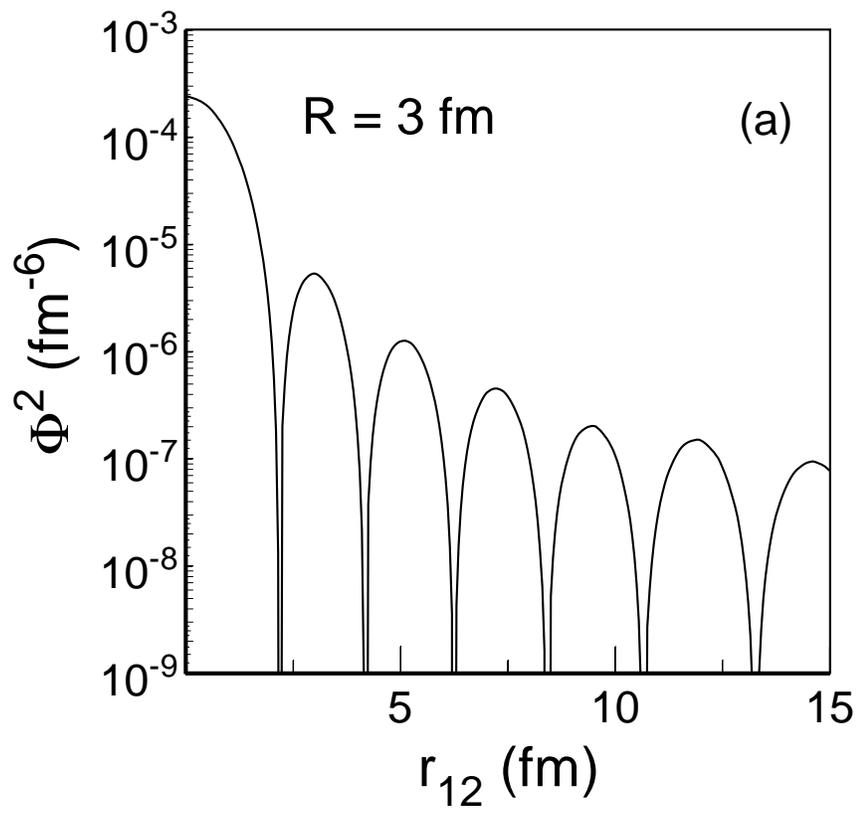
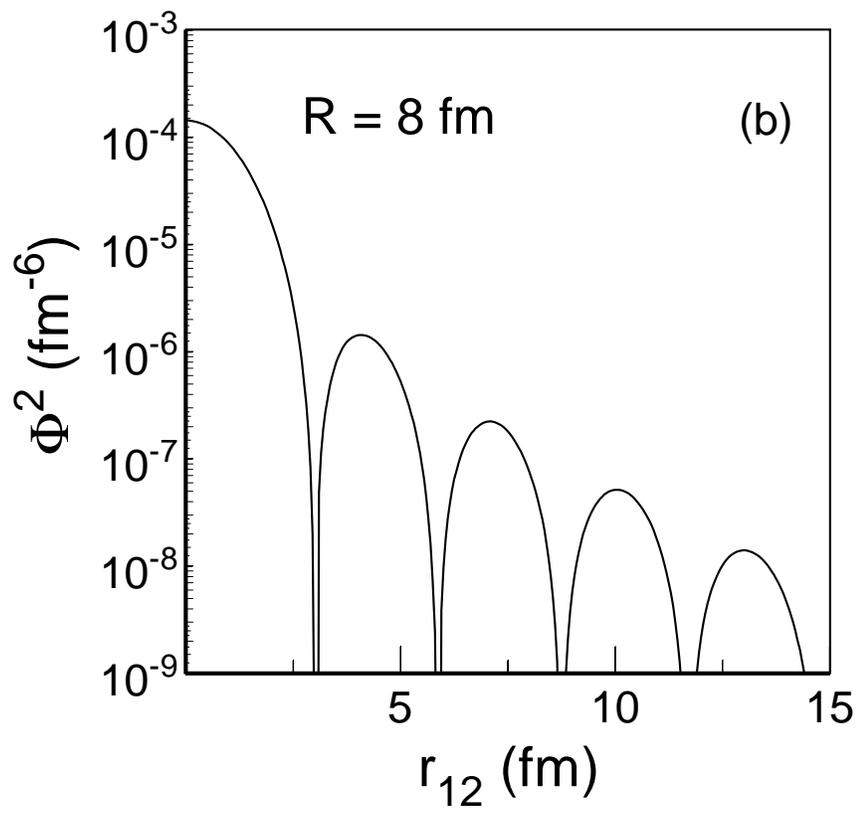
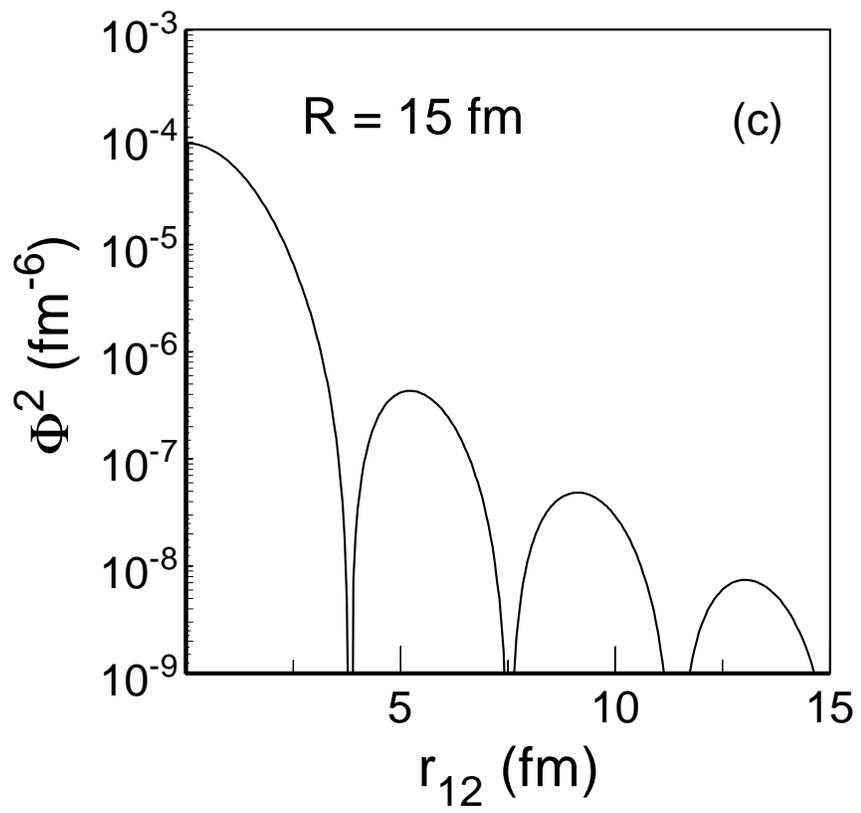
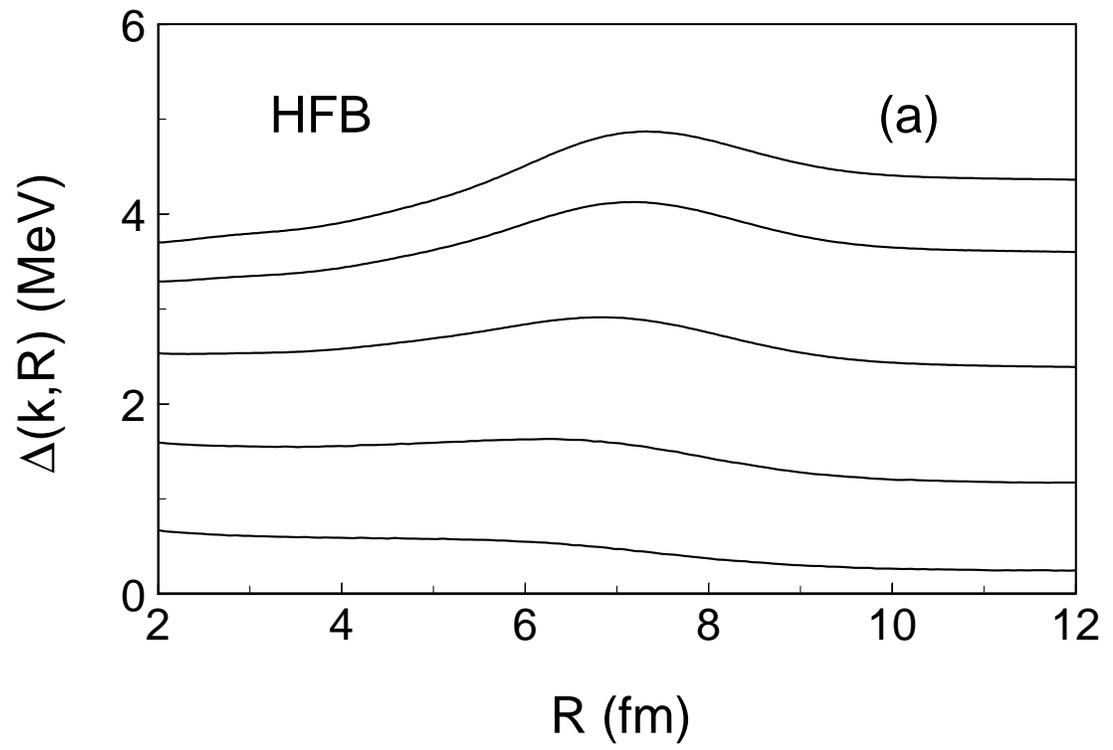
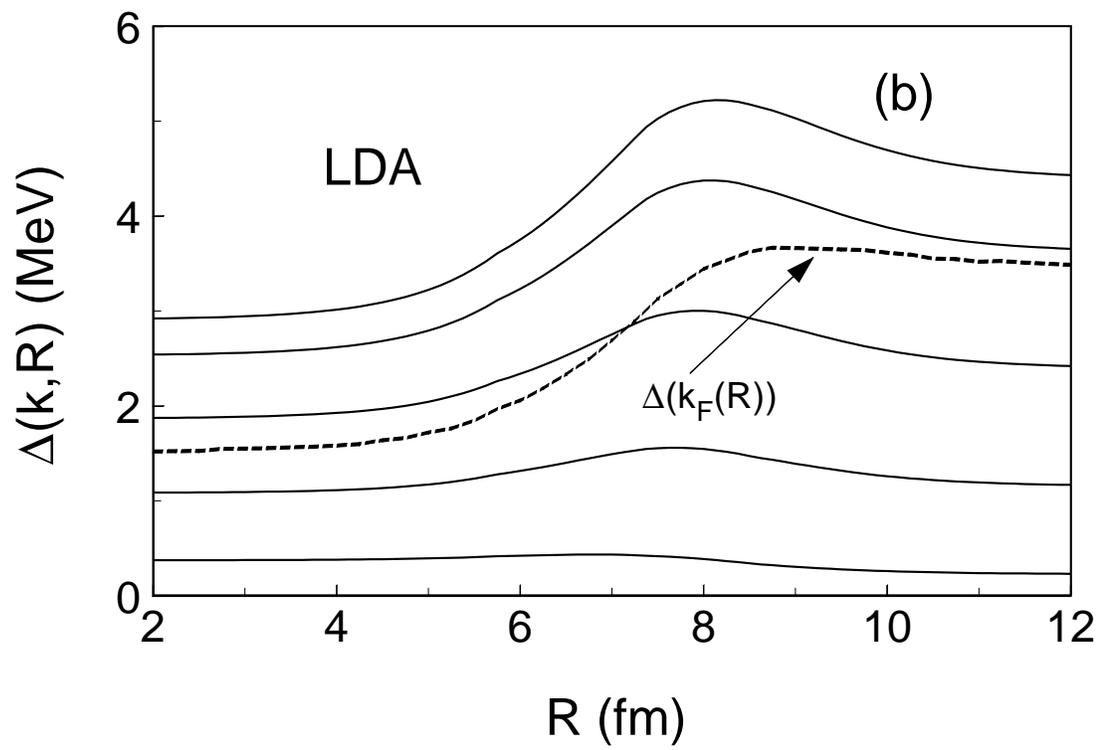
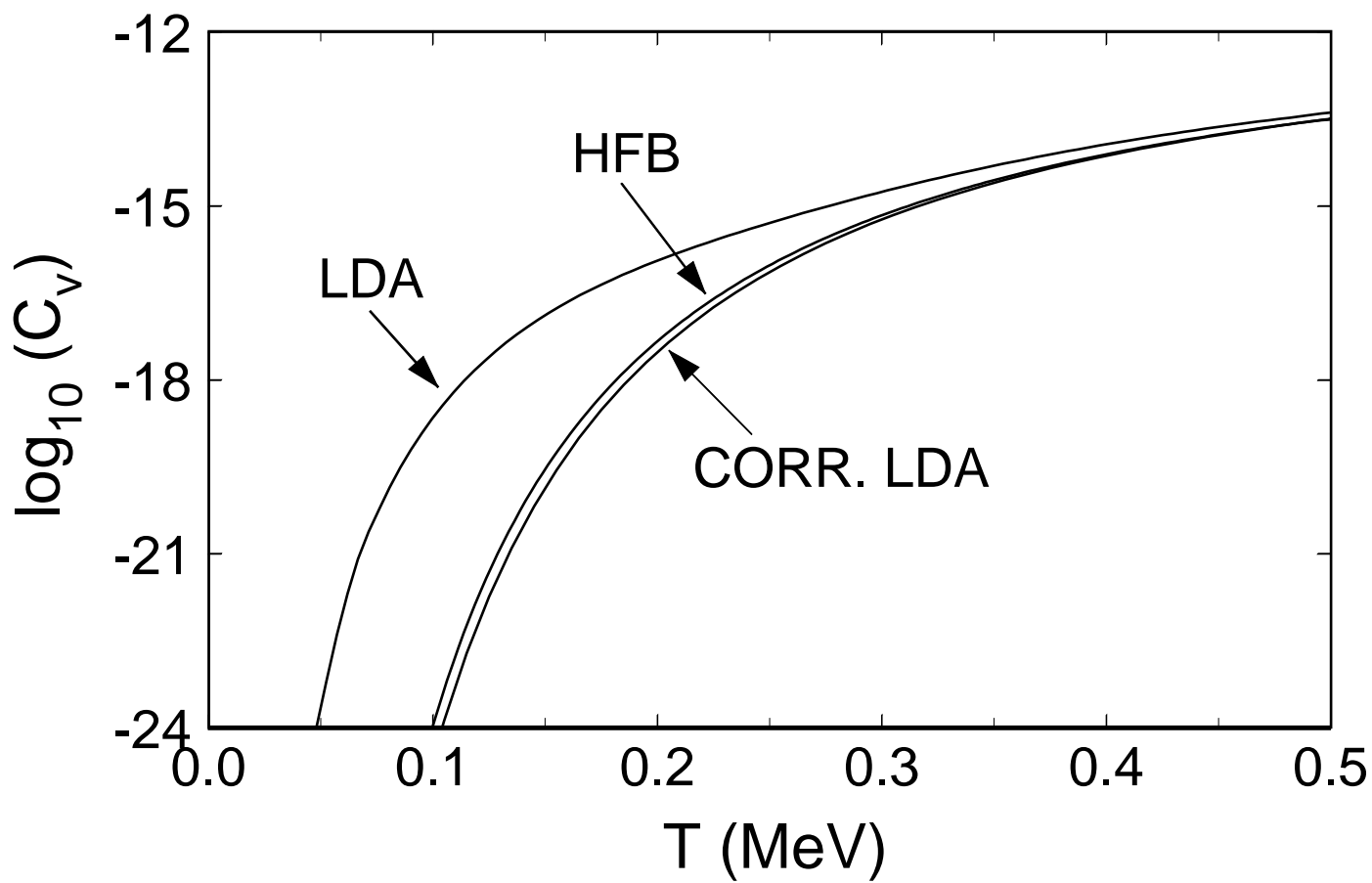
\end{document}